\documentclass{aa}
\usepackage{graphicx}

\begin{document}

\title{The star cluster Collinder~232 in the Carina complex and its
relation to Trumpler~14/16\thanks{Based on observations taken at ESO
La Silla. Tables~ 1 and 2 are available only in electronic form at the CDS via
anonymous ftp to cdsarc.u-strasbg.fr (130.79.128.5) or via http://cdsweb.u-strasbg.fr/cgi-bin/qcat?J/A+A/}}

    \author{ Giovanni Carraro\inst{1,2},
    Martino Romaniello\inst{2},  Paolo Ventura\inst{3}, and Ferdinando 
    Patat\inst{2}}

   \offprints{Giovanni Carraro ({\tt giovanni.carraro@unipd.it})}

   \institute{Dipartimento di Astronomia, Universit\`a di Padova,
	vicolo dell'Osservatorio 2, I-35122, Padova, Italy
     \and
        European Southern Observatory,
 Karl-Schwarzschild-Str 2, D-85748 Garching b. M\"unchen, Germany
    \and
        Osservatorio Astronomico di Roma,
      Via di Frascati 33, I-00040, Monte Porzio Catone, Italy
             }

   \date{Received September 2003; accepted}

\abstract{
In this paper we present and analyze new CCD $UBVRI$ photometry down
to $V~\approx$~21 in the region of the young open cluster
Collinder~232, located in the Carina spiral arm, and discuss its
relationship to Trumpler~14 and Trumpler~16, the two most prominent
young open clusters located in the core of NGC~3372 (the Carina
Nebula).  First of all we study the extinction pattern in the region.
We find that the total to selective absorption ratio $R_V$ differs
from cluster to cluster, being $3.48\pm0.11$, $4.16\pm0.07$ and
$3.73\pm0.01$ for Trumpler~16, Trumpler~14 and Collinder~232,
respectively. Then we derive individual reddenings and intrinsic
colours and magnitudes using the method devised by Romaniello et
al. (2002). Ages, age spreads and distances are then estimated by
comparing the Colour Magnitude Diagrams and the Hertzsprung-Russel
diagram with post and pre-main sequence tracks and isochrones. We find
that Trumpler~14 and Collinder~232 lie at the same distance from the
Sun (about 2.5 kpc), whereas Trumpler~16 lies much further out, at
about 4 kpc from the Sun.  As for the age, we find that Trumpler~16 is
older than both Trumpler~14 and Collinder~232.  For all the clusters
we indicate the existence of a significant age dispersion, whose
precise value is hampered by our inability to properly distinguish
members from non-members. We finally suggest that Collinder~232 is a
physical aggregate and provide estimates of its basic parameters.
}

\authorrunning{Carraro et al.}

\maketitle

%

\section{Introduction}
Aiming at providing a homogeneous photometric database for all the
open clusters located in the Carina complex (Feinstein 1995, Smith et
al 2001), we have carried out an observational program which resulted
in the multicolor $UBVRI$ photometry of 12 star clusters in a $
2^{o}\times 2^{o} $ region around $\eta$~Carin\ae~.  We already
reported on some of these clusters in a series of papers (Carraro et
al 2001, Patat \& Carraro 2001, Carraro \& Patat 2001, Carraro 2002,
Baume et al. 2003).\\ Here we concentrate on Collinder~232 (Collinder
1931) and on the very well studied Trumpler~14 and Trumpler~16
clusters close to $\eta$ Carin\ae~ (Trumpler 1930).  Collinder~232
($\alpha$~=~10:44:48.0, $\delta$~=~ -59:34:00.0, $l$~=~187.51, $b$~=~
-0.54; J2000.0) is located near the northern edge of the Great Carina
Nebula, about $6^{\prime}$ above $\eta$~Carin\ae~.\\ Unlike the other
clusters in this region (Trumpler~14, 15 and 16), which appear rather
compact on sky maps, Collinder~232 is more sparse and less rich in
stars.  Although several observations have been carried out in the
past in this region, a systematic and detailed study of this cluster
is still missing.  Moreover, we analyze the data for one field
centered on Trumpler~14, and 3 fields in the region of Trumpler~16,
aiming at investigating the relationship between Collinder~232 and
these two clusters, in order to establish whether or not they lie at
the same distance from the Sun, whether or not they are coeval, and,
finally, whether or not they are individual objects. These facts, in
turn, are crucial in order to understand the Star Formation (SF)
history of the region.\\
\noindent
These questions have already been addressed many times in the past,
often leading to contradictory results.  A very detailed study of
Trumpler~14 has been conducted by Vazquez et al (1996), whereas a
recent study on Trumpler~16 and Trumpler~14 has been presented by
DeGioia-Eastwood et al (2001), whom the reader is referred to for
further details.  This latter study shows that the two clusters lie at
the same distance and are almost coeval. However the result is
hampered by the assumption that the reddening law is {\it normal} in
the entire region, although previous studies - like for instance
Vazquez et al. (1996) - had convincingly shown that at least in the
region of Trumpler~14 the extinction is anomalous.\\

\noindent
To briefly summarize the current understanding, we follow Walborn
(1995), who provided a nice review of the present status of our
knowledge of the region around $\eta$~Carin\ae~:

\begin{description}
\item $\bullet$ Trumpler ~14  seems to be younger than Trumpler~16;
\item $\bullet$ both cluster lie at the same distance form the Sun;
\item $\bullet$ Collinder~232 is not a physical system, but contains
stars which belong to Trumpler~14 or Trumpler~16;
\item $\bullet$ Collinder~228 is part of Trumpler ~16;
\item $\bullet$ the extinction toward this region is still very
controversial;
\item $\bullet$ If a difference in $R_V$ exists between Trumpler~14 and
16, in the sense that $R_V(Tr14) = R_V(Tr16) + 1$, there would be
no need for either a distance or age difference between the two clusters. 
\end{description}

\noindent

This picture is essentially confirmed by the recent paper by Tapia et
al. (2003).

In this paper we present new $UBVRI$ deep CCD photometry
for all the 3 clusters,
aiming at deriving homogeneous estimates for their fundamental parameters,
like distance, age and interstellar absorption.\\

\begin{figure*}
\centering
\includegraphics[width=16cm,height=16cm]{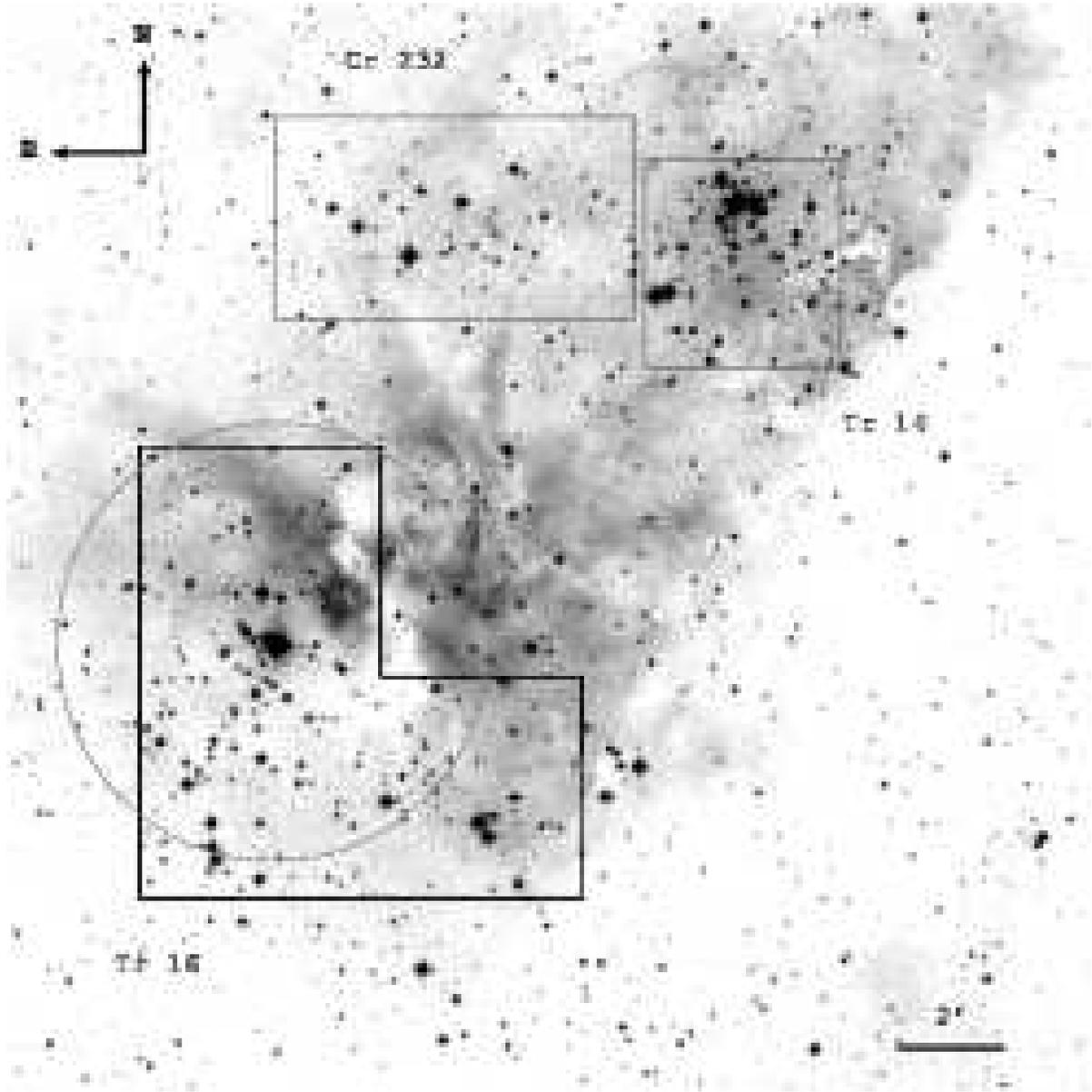}
\caption{A map of the observed regions around Collinder~232,
Trumpler~14 and Trumpler~16. North
is up, East to the left. The field is $20^{\prime} \times 
20^{\prime}$. The circle centered in $\eta$ Carin\ae~ 
has a radius of $4^{\prime}$ and encloses most of the stars
believed to be associated with Trumpler~16. See text for more
details.}
\end{figure*}

The layout of the paper is as follows: Section~2 presents in details
the data acquisition and reduction. In Section~3 we discuss previous
investigations of Collinder~232; in Section~4 we present our data and
compare our photometry with previous ones.  In Section~5 we briefly
summarize Trumpler~14 and 16 properties, and compare our photometry
for these clusters with data available from the literature.  In
Section~6 we critically discuss the extinction pattern in the
direction of the Carina nebula and derive the individual reddening and
membership of stars in Collinder~232, Trumpler~14 and 16.  Section~7
is dedicated to derive estimates for Collinder~232, Trumpler~14 and
Trumpler~16 ages and distances. Then, in Section~8 we discuss the
mutual relationship between the 3 clusters and re-analyse the SF
history in the Carina region, providing the basic conclusions of this
investigation.

\begin{table*}
\tabcolsep 0.30truecm
\caption{Journal of observations of Collinder~232 (April 14, 1996)}
\end{table*}

\begin{table*}
\tabcolsep 0.30truecm
\caption{Journal of observations of Trumpler~14 and Trumpler~16 (April 16, 1996)}
\end{table*}

\section{Observations and Data Reduction}
Observations were conducted at La Silla on April 14-16, 1996, using
the imaging Camera (equipped with a TK coated 512 $\times$ 512 pixels
CCD \#33) mounted at the Cassegrain focus of the 0.92m ESO--Dutch
telescope.  The scale on the chip is 0$^{\prime\prime}.44$ pix$^{-1}$
and the array covers about 3$^\prime$.3 $\times$ 3$^\prime$.3 on the
sky. Due to the projected diameter of the objects and the relatively
small field of view, it was necessary to observe two overlapping
fields for Collinder~232.  The nights were photometric with an average
seeing of 1.6 arcsec.  To allow for a proper photometric calibration
and to asses the night quality, the standard fields RU~149, PG~1657,
SA~109 and SA~110 (Landolt 1992) were monitored each night. Finally, a
series of flat--field frames on the twilight sky were taken. The
scientific exposures have been flat--field and bias corrected by means
of standard routines within {\it IRAF}\footnote{IRAF is distributed by
the National Optical Astronomy Observatories, which is operated by the
Association of Universities for Research in Astronomy, Inc., under
contract to the National Science Foundation.}. Further reductions were
performed using the DAOPHOT-ALLSTAR packages (Stetson 1991) in the
{\it IRAF} environment.  Some details of the observations are given in
the log-book in Table~1 and 2.\\

Moreover in the night of April 16, 1996 we observed 1 field centered
in Trumpler~14, and 3 overlapping fields in the region of
Trumpler~16. The basic information on these observations are reported
in Table~2, whereas the covered regions are shown in Fig.~1, which
reports a DSS image\footnote{Digital Sky Survey,{\tt
http://archive.eso.org/dss/dss}} of a $20^{\prime} \times 20^{\prime}$
region around $\eta$ Carina\ae~.\\

The transformation from instrumental magnitudes to the standard
Kron-Cousins system was obtained with expressions of the form

\begin{equation}\label{eq:mag}
M_i=m_i + zp_i + \gamma_i (M_i-M_j) - k_i z
\end{equation}

where $M_i$, $m_i$, $zp_i$, $\gamma_i$ and $k_i$ are the calibrated
magnitude, instrumental magnitude, zero point, colour term and
extinction coefficient for the $i-th$ passband and $z$ is the
airmass. The transformation requires of course the knowledge of the
reference colour $(M_i-M_j)$, which is easily computed from the
instrumental magnitudes through the following relation:

\begin{equation}\label{eq:col}
(M_i-M_j)=\frac{m_i-m_j + zp_i -zp_j - (k_i-k_j)z}{\gamma_{ij}}
\end{equation}

where we have set $\gamma_{ij}=1-\gamma_i+\gamma_j$. If $\sigma_{mi}$,
$\sigma_{zpi}$, $\sigma_{\gamma i}$ and $\sigma_{ki}$ are the RMS
errors on the instrumental magnitude, zero point, colour term and
extinction coefficient for the $i-th$ passband, formal uncertainties
on calibrated colors are then obtained propagating the various errors
through Eq.~\ref{eq:col} as follows:

\begin{equation}
\sigma_{(Mi-Mj)}^2 \simeq 
\frac{\sigma_{m,ij}^2 + \sigma_{ps,ij}^2 + (M_i-M_j)^2\sigma_{\gamma,ij}^2}{\gamma_{ij}^2}
\end{equation}

For sake of simplicity, we have set
$\sigma_{m,ij}^2=\sigma_{mi}^2+\sigma_{mj}^2$,
$\sigma_{\gamma,ij}^2=\sigma_{\gamma i}^2+\sigma_{\gamma j}^2$ and
$\sigma_{ps,ij}^2 = \sigma_{zp,ij}^2+z^2\sigma_{k,ij}^2$.

Finally, the RMS uncertainties on the calibrated magnitudes are given by:

\begin{equation}
\sigma_{Mi}^2 \simeq
\sigma_{mi}^2 + \sigma_{psi}^2 +  
(M_i-M_j)^2\sigma_{\gamma i}^2 + \gamma_i^2 \sigma_{(Mi-Mj)}^2
\end{equation}

where we have neglected the error on $z$ and assumed that the
images in different passbands have been obtained at very similar 
airmass, as it was in fact the case.

\begin{table}
\caption{\label{tab:photcoeff}Average photometric coefficients
obtained during April 13--16, 1996. ESO--Dutch 0.92m telescope, TK CCD~\#33.}
\begin{tabular}{ccccc}
Filter & Ref. Color & $zp$        & $\gamma$ & $k$ \\
\hline  
 $U$   & $(U-B)$    & 19.85$\pm$0.02 & 0.095$\pm$0.020 & 0.46$\pm$0.02 \\
 $B$   & $(B-V)$    & 21.93$\pm$0.01 & 0.079$\pm$0.010 & 0.27$\pm$0.02 \\
 $V$   & $(B-V)$    & 22.19$\pm$0.01 & 0.030$\pm$0.006 & 0.12$\pm$0.02 \\
 $R$   & $(V-R)$    & 22.18$\pm$0.01 & 0.025$\pm$0.014 & 0.09$\pm$0.02 \\
 $I$   & $(V-I)$    & 21.11$\pm$0.01 & 0.062$\pm$0.006 & 0.06$\pm$0.02 \\
\hline
\end{tabular}
\end{table}

Estimated uncertainties as a function of magnitude are reported in
Tab.~\ref{tab:photerr}, from which it appears clearly that down to
$V\simeq17$ they are dominated by the errors on the photometric
solution, while at fainter magnitudes the contribution by the
poissonian photon shot noise $\sigma_m$ (estimated by DAOPHOT) 
becomes relevant.

\begin{table}
\caption{\label{tab:photerr} Global photometric RMS errors as a function
of magnitude.}
\centerline{
\begin{tabular}{cccccc}
\hline
Mag    &  $\sigma_U$ & $\sigma_B$ & $\sigma_V$ & $\sigma_R$ & $\sigma_I$ \\
\hline 
  9--11 &  0.04 & 0.03 & 0.03 & 0.03 & 0.03 \\
 11--13 &  0.04 & 0.03 & 0.03 & 0.03 & 0.03\\
 13--15 &  0.04 & 0.03 & 0.03 & 0.03 & 0.03\\
 15--17 &  0.05 & 0.03 & 0.03 & 0.03 & 0.03\\
 17--19 &  0.08 & 0.03 & 0.04 & 0.04 & 0.05\\
 19--20 &    -  & 0.04 & 0.05 & 0.07 & 0.09\\
 20--21 &    -  & 0.07 & 0.09 & 0.15 & 0.22\\
 21--22 &    -  & 0.12 & 0.18 & 0.27 &   - \\
\hline
\end{tabular}
}
\end{table}

\begin{figure}
\centering
\includegraphics[width=9cm,height=12cm]{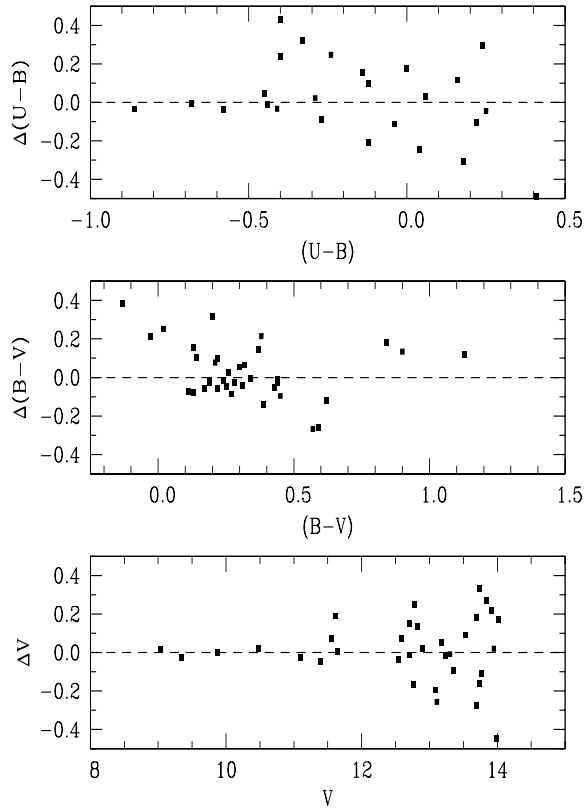}
\caption{A comparison of our photometry with Massey \& Johnson
(1993) study. The comparison is in the sense (this study - Massey \& Johnson).}
\end{figure}

\begin{figure}
\centering
\includegraphics[width=9cm,height=12cm]{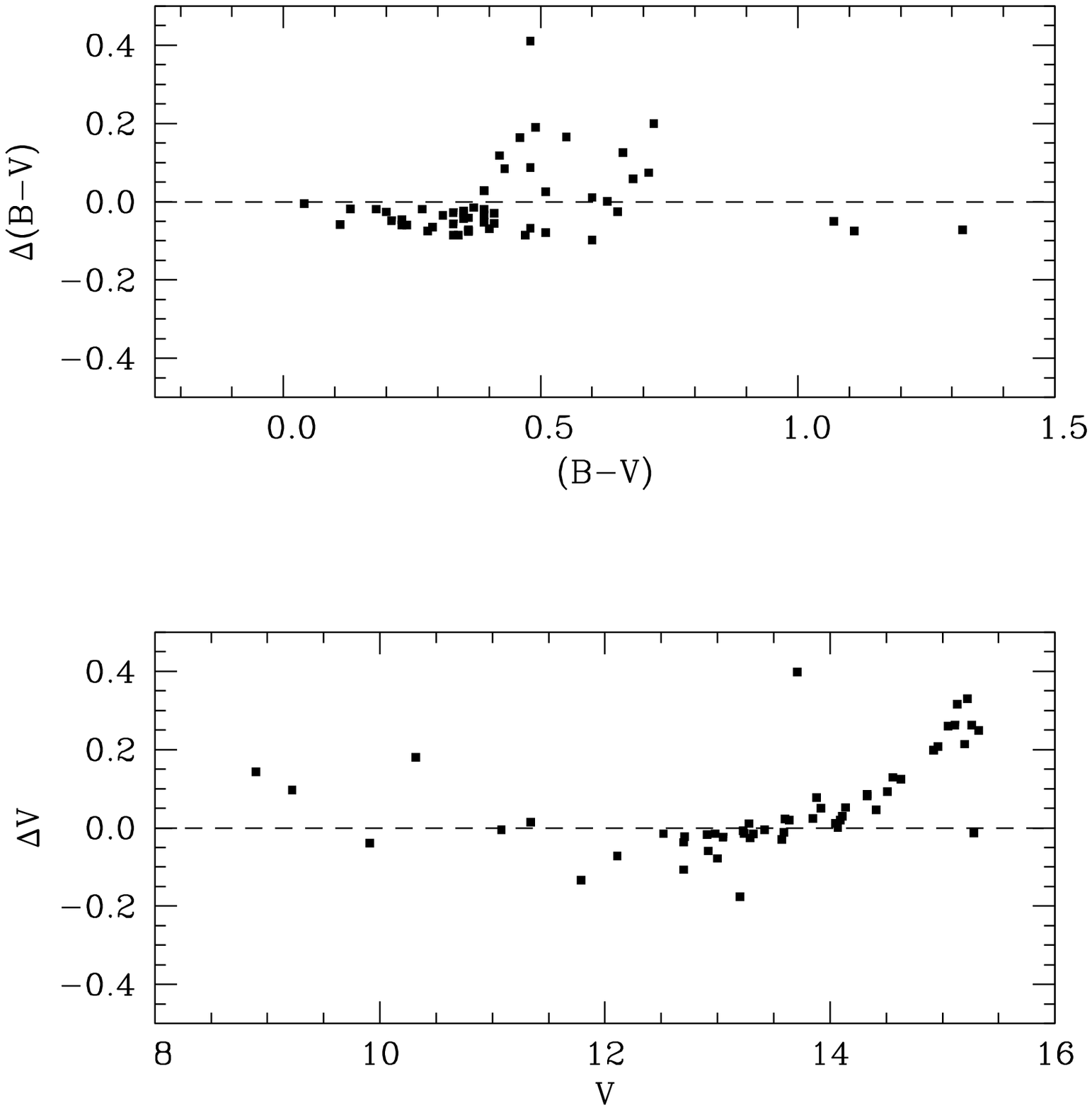}
\caption{A comparison of our photometry with Cudworth et al
(1993) study. The comparison is in the sense (this study - Cudworth et al).}
\end{figure}

\section{Collinder~232: previous results}
Collinder~232 was observed in the past several times due to its
proximity to $\eta$~Carin\ae~ and always in connection with
Trumpler~16.\\ Massey \& Johnson (1993) obtained Schmidt CCD
photometry in $UBV$ bands for about 50 stars down to $V~=~14$ in their
study of the young open clusters Trumpler 14 and 16.\\ Similarly,
Cudworth et al (1993) obtained photographic $BV$ photometry for about
80 stars down to $V~=~15.5$ in the region of Collinder~232 in their
large astrometric survey of star clusters close to $\eta$~Carin\ae~.
Cudworth et al (1993) selected cluster members on the basis of proper
motions, and provided the first Color Magnitude Diagram (CMD) of
Collinder~232, although no estimates are given for the cluster
fundamental parameters.\\ Tapia et al (1988) presented near-infrared
$JHKL$ photometry for 29 stars in Collinder~232. Nonetheless they
associate Collinder~232 with Trumpler~16, and study the inter-stellar
extinction toward these clusters considering them as a single
system.\\ More recently, Tapia et al. (2003) obtained $UBVRIJHK$
photometry in the field of Trumpler~14, 16 and Collinder~232, reaching
approximatively the same limiting magnitude.  Finally, Levato \&
Malaroda (1982) provide spectral classification for 4 stars in the
field of Collinder~232.

\section{The present study}
We provide $UBVRI$ photometry for 970 stars in a $6^{\prime}.3 \times
3^{\prime}.5$ region centered in Collinder~232.  Limiting magnitudes
($5 \sigma$) are $U~=~17$, $B~=~22.3$, $V~=~21.6$,
$R~=~20.9$ and $I~=~20.6$.  The region we sampled is shown in Fig.~1,
where a $V$ map is presented. In this map North is on the top, East to
the left. Fig.~2 and Fig.~3 show the comparison of our photometry with
the one of Massey \& Johnson (40 common stars) and Cudworth et al (60
common stars), respectively.\\ In the case of Fig.~2 we notice that
the agreement in magnitude is good up to $V~=~12.0$, and below there
is a large scatter.  In the case of colour the same scatter is
present, but there is no systematic difference.  We interpret the
large scatter as due to Massey \& Johnson (1993) photometry, which was
obtained with a small Schimdt telescope and a CCD having a very large
scale, almost $2^{\prime\prime}$/pixels. Although the field is not
particularly crowded, some stars are actually blended.  Finally, the
typical error at $V~=~13.0-14.0$ is in the range 0.05-0.10 mag in the
Massey \& Johnson (1993) photometry, while in our case is 0.02-0.04
(Patat \& Carraro 2001). By considering all the stars, we get\\

\[
V_{CRVP} - V_{MJ} = 0.125\pm0.477
\]

\[
(B-V)_{CRVP} - (B-V)_{MJ} = 0.027\pm0.143 
\]

\[
(U-B)_{CRVP} - (U-B)_{MJ} = -0.126\pm0.424 
\]

\noindent
where the suffix $CRVP$ refers to this study,  and $MJ$
to Massey \& Johnson (1993). These numbers mirror the results of
Fig.~2, emphasizing the existence of a large scatter. We stress
however, that for $V$ brighter than 11.5, the two photometries are
consistent.\\

Some scatter is also visible in the comparison with Cudworth et al
(1993) photographic photometry (Fig.~3). In this case the major source
of errors is the poor precision of photographic photometry at the
faint magnitude end, and the poor treatment of star blending in
crowded regions. By considering only the stars brighter than $V~=~14$
we get

\[
V_{CRVP} - V_{CMDE} = 0.004\pm0.100
\]

\[
(B-V)_{CRVP} - (B-V)_{CMDE} = -0.049\pm0.024 
\]

\noindent
which means that the two photometries are consistent up to this magnitude,
and then the deviation becomes very large.

\begin{figure}
\centering
\includegraphics[width=9cm,height=9cm]{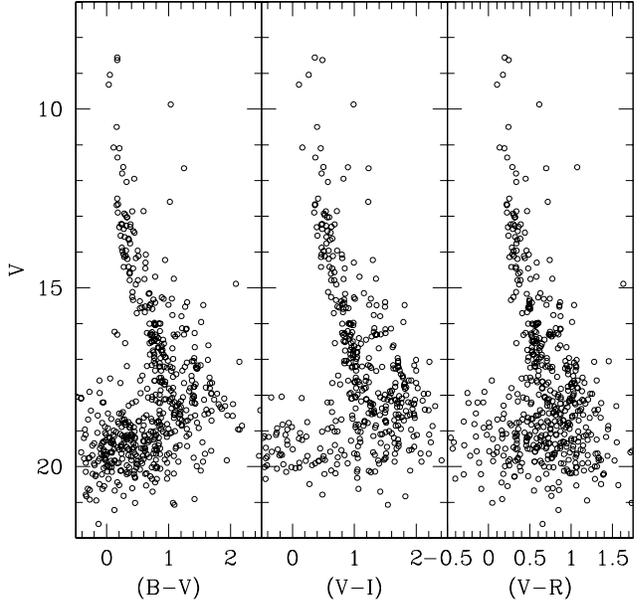}
 \caption{The CMDs of Collinder~232 including all the detected stars.}
\label{cmd232}
\end{figure}

The CMDs from our photometry for all the measured stars is plotted in
Fig.~4 in the planes $V-(B-V)$, $V-(V-I)$ and $V-(V-R)$. 
Our photometry reaches $V\approx 21$, although below
$V\approx 18$ the scatter in color is quite large.  This is mostly due
to background star contamination, and only partially to photometric
errors and the presence of unresolved binary systems, whose percentage
in these clusters is around 30$\%$ (Levato et al 1990).\\

As for data completeness (V magnitude), 
we performed an analysis by using IRAF tasks 
ADDSTAR, which yields 100$\%$ down to V=17.0, 93$\%$ down to V=18.2
and  57$\%$ down to V=19.0 .

\begin{figure}
\centering
\includegraphics[width=9cm,height=9cm]{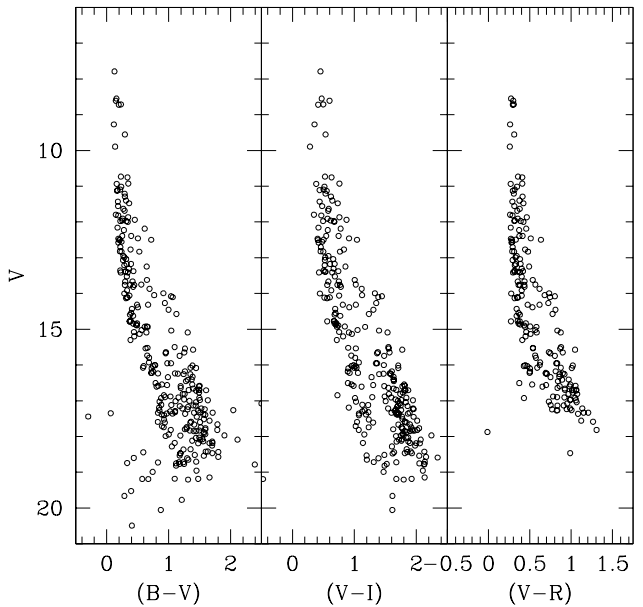}   
\caption{The CMDs of Trumpler~14 including all the detected stars.}
\end{figure}

\begin{figure}
\centering
\includegraphics[width=9cm,height=9cm]{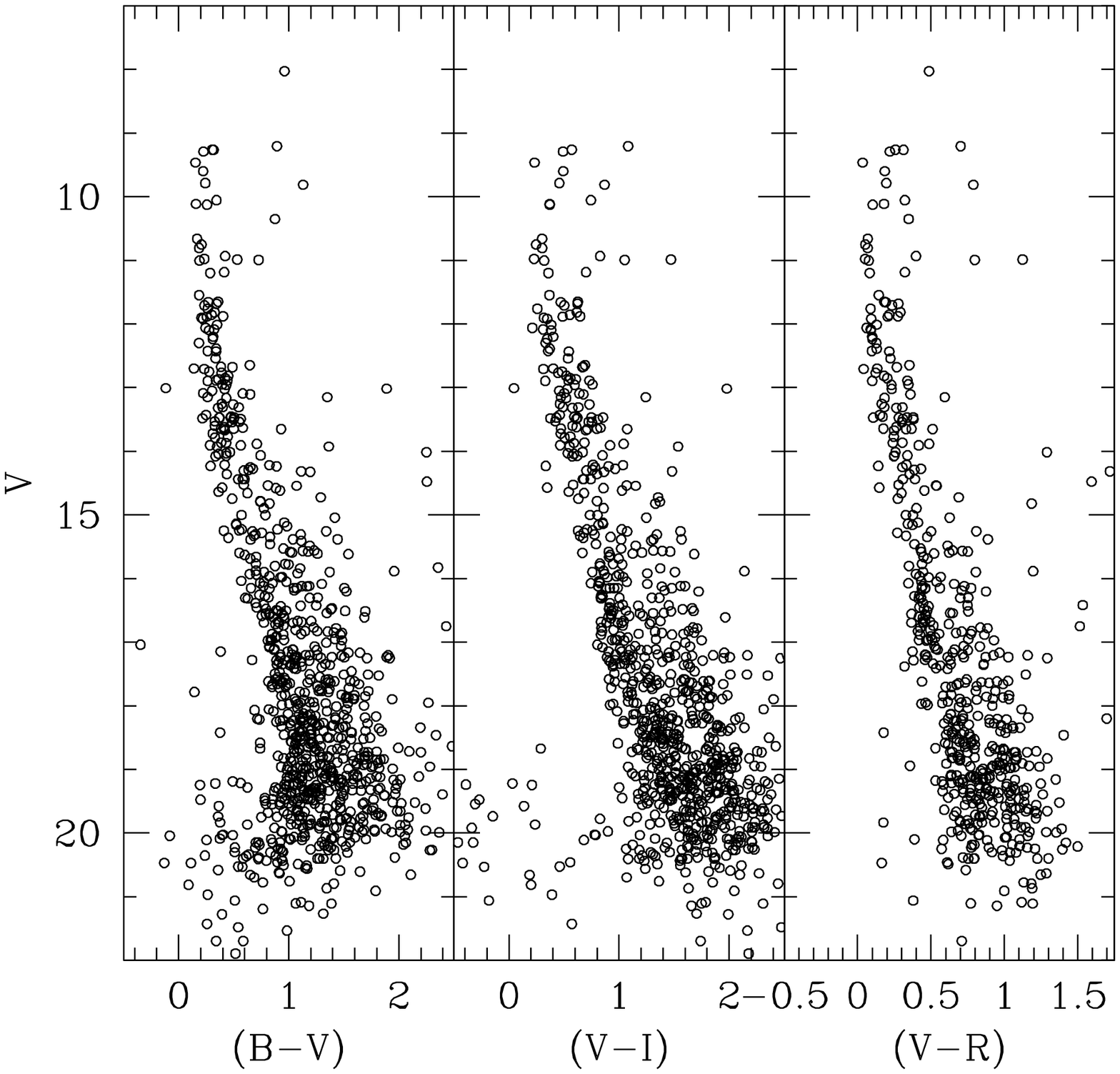}   
\caption{The CMDs of Trumpler~16 including all the detected stars.}
\end{figure}

\begin{table*}
\tabcolsep 0.50cm
\caption{Extinction parameters of Trumpler~14 stars with near IR
photometry and spectral classification.}
\begin{tabular}{ccccccccc}
\hline
\multicolumn{1}{c}{ID} &
\multicolumn{1}{c}{FMM73.} &
\multicolumn{1}{c}{$Sp.~Type$} &
\multicolumn{1}{c}{$E(B-V)$}  &
\multicolumn{1}{c}{$R_V (i)$} &
\multicolumn{1}{c}{$R_V (iii)$} & 
\multicolumn{1}{c}{$A_V(ii)$}  &
\multicolumn{1}{c}{$R_V(ii)$} \\
\\
\hline   
   7&  20&  O6-V   &  0.596& 3.67& 3.46& 1.79$\pm$0.05&  3.00$\pm$0.13&\\ 
  10&  21&  09-V   &  0.618& 3.96& 3.83& 2.03$\pm$0.13&  3.28$\pm$0.16&\\ 
  16&  27&  B1-V   &  0.521& 4.45& 4.51& 2.05$\pm$0.21&  3.94$\pm$0.29&\\ 
  27&  18&  B0-V   &  0.520& 4.76& 4.26& 2.18$\pm$0.33&  4.19$\pm$0.41&\\ 
  29&  15&  B7-V   &  0.477& 4.56& 4.42& 1.91$\pm$0.41&  4.00$\pm$0.48&\\ 
  31&  26&  B2-V   &  0.529& 4.85& 4.49& 2.16$\pm$0.09&  4.08$\pm$0.17&\\
  32&  23&  B1-V   &  0.571& 4.24& 3.79& 1.96$\pm$0.24&  3.44$\pm$0.37&\\  
  35&  22&  B2-V   &  0.489& 4.65& 4.24& 1.90$\pm$0.07&  3.88$\pm$0.15&\\ 
  36&  28&  B2-V   &  0.606& 4.23& 2.90& 2.17$\pm$0.13&  3.59$\pm$0.22&\\ 
  42&  12&  B2-V   &  0.398& 5.38& 4.87& 1.87$\pm$0.19&  4.70$\pm$0.25&\\ 
\hline
\end{tabular}
\end{table*}

\begin{table*}
\tabcolsep 0.50cm
\caption{Same as in Table~5, but for Collinder~232.}
\begin{tabular}{cccccccc}
\hline
\multicolumn{1}{c}{ID} &
\multicolumn{1}{c}{FMM73.} &
\multicolumn{1}{c}{$Sp.~Type$} &
\multicolumn{1}{c}{$E(B-V)$}  &
\multicolumn{1}{c}{$R_V (i)$} &
\multicolumn{1}{c}{$R_V (iii)$} & 
\multicolumn{1}{c}{$A_V(ii)$}  &
\multicolumn{1}{c}{$R_V(ii)$} \\
\hline   
   1&  HD~93160&  O6.5-V  &  0.472& 3.82& 3.68 & 1.55$\pm$0.05&  3.29$\pm$0.13\\ 
   2&  HD~93161&  06-III  &  0.470& 4.12& 3.90 & 1.63$\pm$0.13&  3.47$\pm$0.16\\ 
   6&  31      &  B0-V    &  0.421& 4.44& 3.88 & 1.52$\pm$0.21&  3.62$\pm$0.29\\ 
\hline
\end{tabular}
\end{table*}

\begin{table*}
\tabcolsep 0.50cm
\caption{Same as in Table~5, but for Trumpler~16.}
\begin{tabular}{ccccccccc}
\hline
\multicolumn{1}{c}{ID} &
\multicolumn{1}{c}{FMM73.} &
\multicolumn{1}{c}{$Sp.~Type$} &
\multicolumn{1}{c}{$E(B-V)$}  &
\multicolumn{1}{c}{$R_V (i)$} &
\multicolumn{1}{c}{$R_V (iii)$} & 
\multicolumn{1}{c}{$A_V(ii)$}  &
\multicolumn{1}{c}{$R_V(ii)$} \\
\hline   
   2& 110      &  O7-V    &  0.606& 4.35 & 4.10 &2.22$\pm$0.05&  3.67$\pm$0.13&\\ 
   3&  34      &  O8.5-V  &  0.610& 3.75 & 3.66 &2.06$\pm$0.13&  3.37$\pm$0.16&\\ 
   4&  27      &  O4.5-V  &  0.606& 4.19 & 4.07 &2.26$\pm$0.21&  3.73$\pm$0.29&\\ 
   5&   1      &  O9.5-V  &  0.422& 1.68 & 0.89 &0.46$\pm$0.33&  1.09$\pm$0.41&\\ 
   7&  HD~93343&  O8-V    &  0.508& 4.37 & 4.10 &1.97$\pm$0.41&  3.89$\pm$0.48&\\ 
   8&   9      &  O9.5-V  &  0.512& 3.84 & 3.58 &1.66$\pm$0.09&  3.38$\pm$0.17&\\
   9&  23      &  O7-V    &  0.633& 2.74 & 2.42 &3.44$\pm$0.24&  2.63$\pm$0.37&\\  
  12&   3      &  O9-V    &  0.537& 2.94 & 2.43 &1.47$\pm$0.07&  2.74$\pm$0.15&\\ 
  15&   8      &  B1.5-V  &  0.427& 3.54 & 3.22 &1.50$\pm$0.13&  3.51$\pm$0.22&\\ 
  16&   2      &  B1.5-V  &  0.406& 3.01 & 2.63 &1.06$\pm$0.19&  2.61$\pm$0.25&\\ 
  20&  65      &  B1.5-V  &  0.409& 4.56 & 4.24 &1.59$\pm$0.07&  3.89$\pm$0.15&\\ 
  22&  22      &  O8.5-V  &  0.707& 3.01 & 3.00 &1.87$\pm$0.13&  2.65$\pm$0.22&\\ 
  24&   4      &  B2-V    &  0.497& 4.61 & 4.38 &1.99$\pm$0.19&  4.00$\pm$0.25&\\ 
  25&  12      &  B2-V    &  0.624& 3.91 & 4.01 &2.25$\pm$0.19&  3.60$\pm$0.25&\\ 
\hline
\end{tabular}
\end{table*}

\section{Previous results for Trumpler~14 and 16}
We report here photometry of a field centered on Trumpler~14, and 3
overlapping fields in the region of Trumpler~16 (see Fig.~1).  In the case
of Trumpler~16 the limiting magnitudes are $U~=~19.9$, $B~=~21.0$,$V~=~20.6$,
$R~=~20.1$ and $I~=~19.9$, whereas for Teumpler~14
the limiting magnitudes are $U~=~19.9$, $B~=~20.1$,$V~=~19.1$,
$R~=~18.0$ and $I~=~20.5$.  As for Collinder~232, we
compare our photometry with previous ones. Since recent studies
usually provide a comparison with the photoelectric photometry by
Feinstein et (1973), we report here the same comparison. Based on
this, comparisons with other photometric studies can be quickly
performed.\\

For Trumpler~14 (27 stars in common), we obtain

\[
V_{CRVP} - V_{FFM} = -0.06\pm0.16
\]

\[
(B-V)_{CRVP} - (B-V)_{FFM} = -0.02\pm0.100 
\]

\[
(U-B)_{CRVP} - (U-B)_{FFM} = -0.06\pm0.11 
\]

\noindent
whereas for Trumpler~16 (44 common stars), we obtain

\[
V_{CRVP} - V_{FFM} = -0.04\pm0.12
\]

\[
(B-V)_{CRVP} - (B-V)_{FFM} = -0.04\pm0.04 
\]

\[
(U-B)_{CRVP} - (U-B)_{FFM} = -0.06\pm0.13 
\]

\noindent
where the suffix $FFM$ refers to Feinstein et al (1973) photometry.\\
The CMDs for all the measured stars are plotted in Fig.~5 and Fig.~6
for Trumpler~14 (343 stars) and Trumpler~16 (1100 stars),
respectively.  In the case of Trumpler~14, our photometry reaches
$V\approx 19$, and is as deep as that presented by Vazquez et
al. (1996).  As for Trumpler~16, our photometry reaches $V\approx 21$,
two magnitudes deeper that the study of DeGioia-Eastwood et al. (2001)
The same kind of comments as for Collinder~232 CMDs can be done both
for Trumpler~14 and 16.  The Main Sequence (MS) is well defined for
almost all its extension, but below $V \approx 17-18$ the scatter in
color is quite large.  This is mostly due to background star
contamination, and only partially to photometric errors and the
presence of unresolved binary systems, whose percentage in these
clusters is also around 30$\%$ (Levato et al 1990).\\

In the case of Trumpler~16, the data completeness (V magnitude) 
analysis provides 100$\%$ down to V=17.4, 91$\%$ down to V=18.6
and  57$\%$ down to V=19.3 . As for Trumpler~14, we find
100$\%$ down to V=16.9, 89$\%$ down to V=17.4
and  51$\%$ down to V=18.1 .

\section{The interstellar extinction toward Collinder~232, Trumpler~14 and
Trumpler~16}

The Carina region is characterized by a remarkable concentration of
young stars (it contains a sizable fraction of the known OB stars of
the Galaxy; Walborn, 1995; Tapia et al. 2003), and gas (the large HII
region NGC~3372).  As shown in Figure~1, the interstellar medium
appears to be very clumpy and great care should be taken to treat the
interstellar extinction properly, as it is not possible to adopt a
unique average extinction law over the whole region (see also Th\'e \&
Graafland, 1995).  Tapia et al. (2003) do not perform a new analysis
of the problem, but simply adopt previous findings by Tapia et
al. (1988), Smith (1987)  and results
presented at the 1995 La Plata workshop on Carina (1995, 
Rev. Mex. Astron Astrof. 2).\\

Here we analyse this issue in a completely new fashion, and tackle the
problem of variable extinction by first deriving the appropriate law
for each cluster, and then applying it to deredden the individual
stars in each of them.

\subsection{The reddening laws}

In order to estimate the selective extinction $R_V=A_V/E(B-V)$ toward
each cluster, we combine our UV-optical $UBVRI$ photometry with the
near-IR one in the $JHKL$ bands from Tapia et al (1988) and the
spectral classification from Levato \& Malaroda (1982) and Morrell et
al (1988). The comparison between the measured and intrinsic colours
expected for the stars' spectral type (Wegner 1994) allows one to
compute the colour excesses in the different bands and, ultimately,
$R_V$. To do so, we have applied three different methods and compared
the results:

\noindent
(i) The first method is based on the following approximate relation
(\emph{e.g.} Whittet, 1992):

\begin{equation}
R_V \simeq 1.1\times\frac{E(V-K)}{E(B-V)}
\end{equation}
which relies on the fact that, as the wavelength increases, the
reddening law becomes less dependent on the nature of the dust grains
and, hence, the ratio between absorption in the $V$ and $K$ band is
nearly a constant along different lines of sight. Its weak point,
though, is that it only uses the flux in two spectral regions, and not
the entire extinction curve.

(ii) The second method we employ partially overcomes this limitation
by using the extinction curve redwards of the R band, but requires its
shape to be known a priori (Morbidelli et al, 1997; Patriarchi et al
2001; see also Carraro, 2002 for an application to Trumpler~15).

First, $A_V$ is determined with a least-square fit to the following
relation:

\begin{equation}
E(\lambda -V) = A_V \times (R_L(\lambda) - 1)
\label{morbi}
\end{equation}
where $\lambda = R,I,J,K,L$ and $R_L$ is the extinction curve
$A_\lambda/A_V$. We have adopted the one from Rieke \& Lebofsky
(1985). Then, $R_V$ is computed from the measured $E(B-V)$ and the
value of $A_V$ derived above.

Since the fitting equation~\ref{morbi} is a homogeneous one, the
uncertainty on $A_V$ for each input star has been computed by
considering $N-1$ degrees of freedom, $N$ being the number of
photometric bands available. This implies that we can obtain only a
lower limit on the $R_V$ uncertainty, since it is rather difficult to
take into account spectral mis-classification and, hence, inaccuracy
in the adopted intrinsic colors (Patriarchi et al 2001).

(iii) In the third and last method the information in all the
available passbands is used. In addition, no assumptions are made on
the extinction law, but, rather, $R_V$ is derived by extrapolating the
extinction curve to infinite wavelengths:

\begin{equation}
R_V=\lim_{1/\lambda\rightarrow 0} E(V-\lambda)/E(B-V)
\end{equation}
under the (obvious) assumption that:

\begin{equation}
\lim_{1/\lambda\rightarrow 0} A(\lambda)=0
\end{equation}

In practice, the measured values of $E(V-\lambda)/E(B-V)$ in the
available bands are fitted with a $5^{\mathrm{th}}$ order polynomial,
which is then extrapolated to $1/\lambda=0$. The linear term in the
polynomial is set to 0 to ensure that the extrapolated curve is
horizontal at the limit.

In addition to $R_V$ this method also yields the complete reddening
curve.  While $R_V$ differs significantly from star to star (see
Tables 5, 6 and 7 for Trumpler~14, Collinder~232 and Trumpler~16,
respectively), the \emph{shape} of the reddening curve shows, within
the accuracy of our measurements, no significant variations and
follows closely the one by Riecke \& Lebofsky (1985). Therefore, in
the following we will adopt it, scaled to the appropriate values of
$R_V$, to deredden our target stars.

The results for $R_V$ are summarized in Tables~5, 6 and 7 for
Trumpler~14 (10 stars), Collinder~232 (3 stars) and Trumpler~16 (14
stars), respectively. There, the stars' identification is reported
together with the individual reddening E$(B-V)$, the value of $R_V$ as
obtained with methods {\it (i)}, {\it (ii)} and {\it (iii)} and the
total absorption $A_V$ derived from the method {\it (ii)}.

The mean $A_V$ for Trumpler~14 and 16 turn out to be 2.0$\pm$0.13 and
1.84$\pm$0.65, and are consistent with the values reported by Tapia et
al. (2003).\\ Even by a cursory inspection of these Tables, it is
clear that there are large variations in $R_V$ not only from cluster
to cluster, but also from star to star within the same cluster.\\

In Table~8, we finally report for each cluster the adopted mean value
of $R_V$ from the three different methods, as obtained by performing
an arithmetic mean trough the data listed in Table~5 to 7.

Table~8 is very useful to compare the $R_V$ values obtained from
different methods.  It appears that the methods {\it (i)} and {\it
(iii)} produce comparable results, whereas method {\it (ii)} has a
tendency to provide lower values of $R_V$.  The only case for which
all the 3 methods yield the same result is the case of Trumpler~16,
for which the number of stars is the largest one.  This immediately
raises the suspect that all methods probably would yield comparable
results, when a sufficient number of stars were available.  Obviously,
this hypothesis needs to be validated.

Nonetheless, since all these three methods have {\it pro} and {\it
contra}, we opted for the adoption of individual cluster $R_Vs$
estimated by extracting a weighted mean of the three methods. These
values are reported in the last column of Table~8 together with the
weighted errors. In the case of Collinder~232 the reported error is
artificially small, being the statistics very poor.\\

Trumpler~14 has the largest value of $R_V$ and Trumpler~16 the lowest
one, with Collinder~232 in the middle, both in the mean value and in
the individual determinations.

The value $R_V=4.16\pm0.07$ we obtain for Trumpler~14 is in good
agreement with the one found by Vazquez et al. (1996) using a variety
of methods. Finally, we note that the relation we find between
Trumpler~14 and 16,
\emph{i.e.} $R_V(Tr14)=R_V(Tr16)+0.68(\pm0.20)$, is only 
in marginal agreement with the one by Th\'e \& Graafland (1995). \\

Regrettably, the paucity of data available for Collinder~232 does not
allow one to draw any firm conclusions on the behaviour of the dust in
it. What we can say, however, is that $R_V$ is definitely different
from Trumpler~14 and 16, independently of the claim that Collinder~232
is not a real cluster, but, rather, its stars belong to either of its
two neighbors.

\begin{table}
\tabcolsep 0.10cm
\caption{Estimates of $R_V$ for the clusters under study.}
\begin{tabular}{cccccc}
\hline
\multicolumn{1}{c}{Cluster} &
\multicolumn{1}{c}{$R_V (i)$} &
\multicolumn{1}{c}{$R_V (iii)$} & 
\multicolumn{1}{c}{$R_V(ii)$} &
\multicolumn{1}{c}{Adopted} \\
\hline   
Trumpler~16   &  3.69$\pm$0.55 & 3.45$\pm$0.69 & 3.31$\pm$0.51 & 3.48$\pm$0.33\\ 
Trumpler~14   &  4.47$\pm$0.48 & 4.18$\pm$0.42 & 3.81$\pm$0.49 & 4.16$\pm$0.21\\ 
Collinder~232 &  4.13$\pm$0.25 & 3.82$\pm$0.10 & 3.46$\pm$0.14 & 3.73$\pm$0.03\\ 
\hline
\end{tabular}
\end{table}

\subsection{Individual reddenings and membership}

In the previous section we have determined the appropriate extinction
curve for every cluster, \emph{i.e.} the one from Rieke \& Lebofsky
normalized to the values of $R_V$ listed in the last column of
Table~8. We can now use it, together with our $UBVRI$ photometry, to
deredden all the stars we have detected in the three clusters. To do
so, we have applied the technique developed by Romaniello et al
(2002). In brief, given a reddening curve and a set of stellar
atmosphere models (the ones by Bessel et al 1998, in our case), the
extinction coefficients and intrinsic magnitudes are computed as a
function of the effective temperature (and, in the case of the
absorption coefficients, also optical depth). The models are, then,
reddened by different amounts of E(B-V) and a $\chi^2$ technique is
applied to determine the best combination of $T_{eff}$ and $E(B-V)$
for every star.  The results are discussed in details in the next
Section.

\begin{figure*}
\centering
\includegraphics[width=16cm,height=16cm]{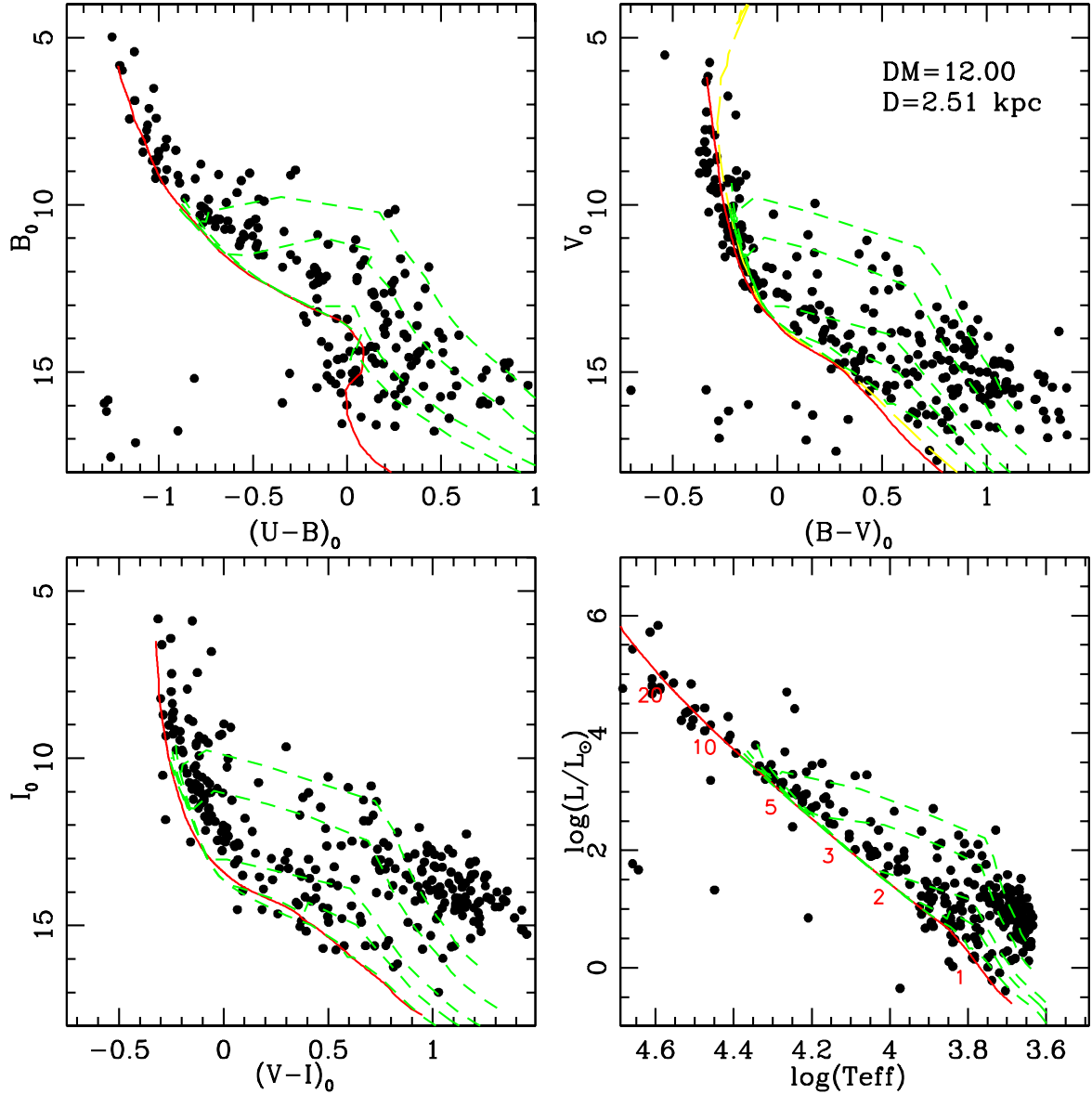}   
\caption{CMDs and HR diagram for the stars in the field of
Trumpler~14. Dashed lines are pre-MS isochrones from Ventura et
al. (1998) for the ages of 0.5, 1, 5, 10 an 20 million years. The
solid line is an empirical ZAMS. In the HR diagram the number along
the sequence indicate the star masses.}
\end{figure*}

\section{Clusters parameters}

\begin{figure*}
\centering
\includegraphics[width=16cm,height=16cm]{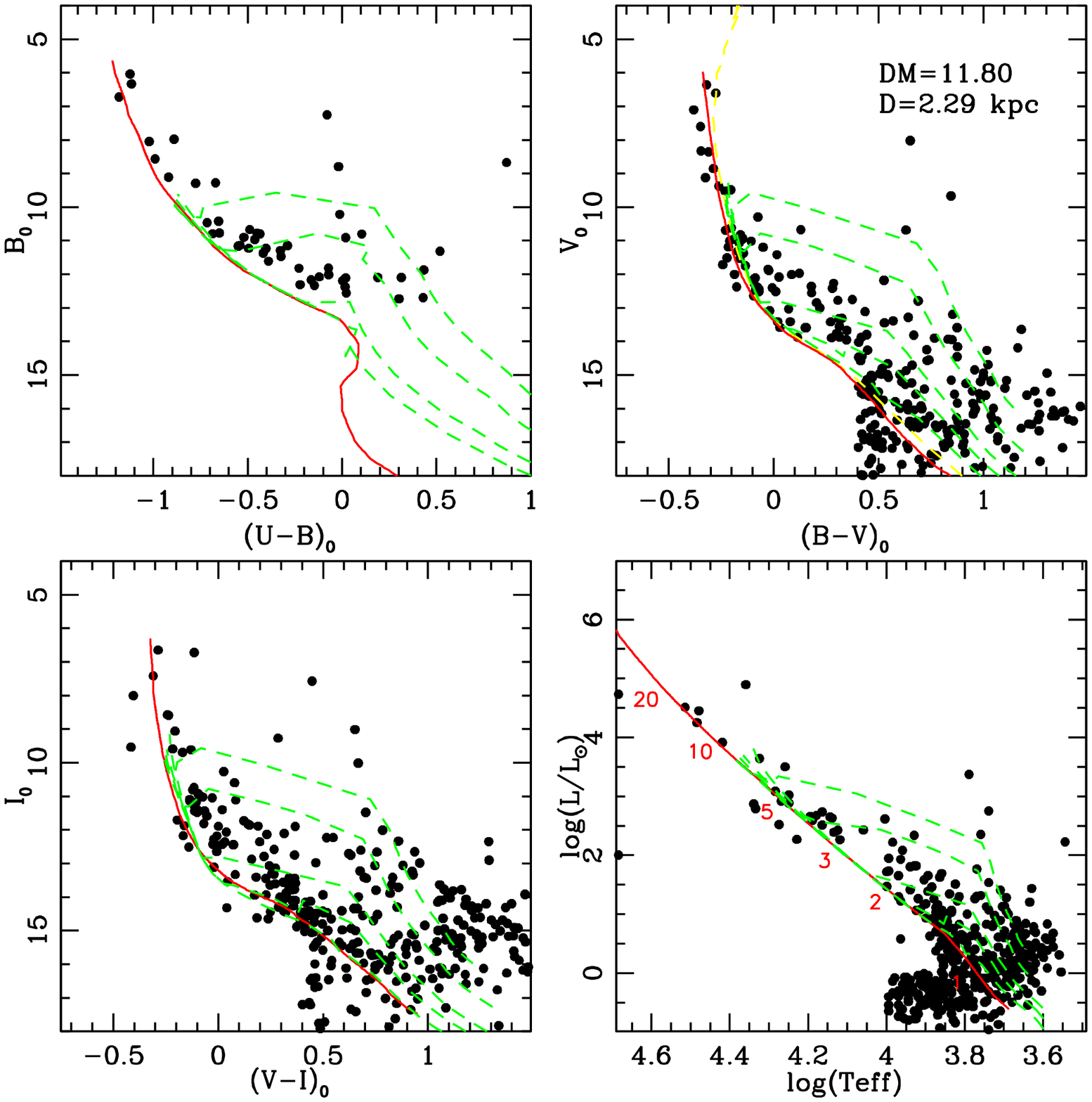}   
\caption{CMDs and HR diagram for the stars in the field of
Collinder~232. Dashed lines are pre-MS isochrones from Ventura et
al. (1998) for the ages of 0.5, 1, 5, 10 an 20 million years. The
solid line is an empirical ZAMS. In the HR diagram the number along
the sequence indicate the star masses.}
\end{figure*}

\begin{figure*}
\centering
\includegraphics[width=16cm,height=16cm]{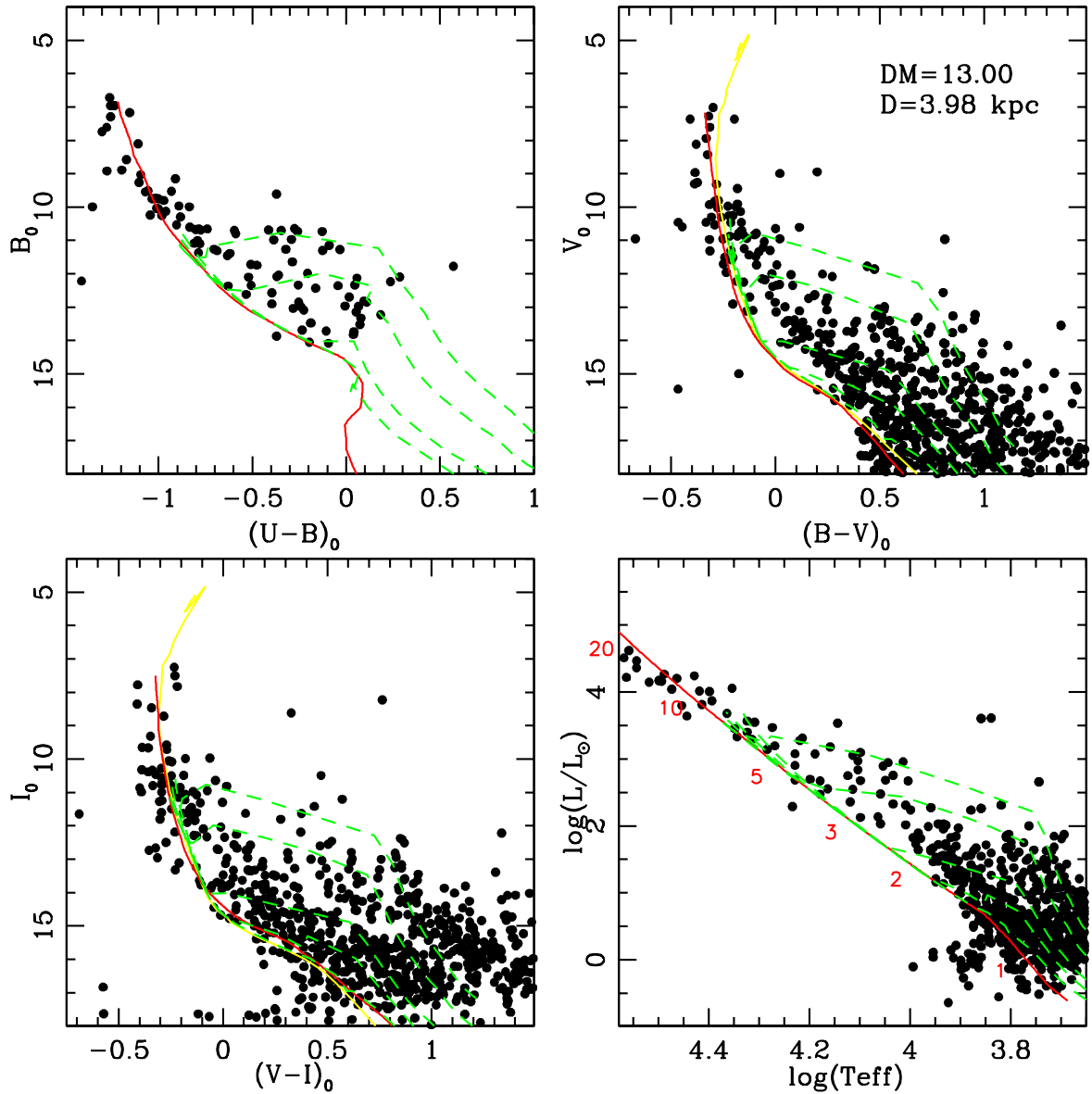}   
\caption{CMDs and HR diagram for the stars in the field of
Trumpler~16. Dashed lines are pre-MS isochrones from Ventura et
al. (1998) for the ages of 0.5, 1, 5, 10 an 20 million years. The
solid line is the empirical ZAMS shifted by 
(m-M) = 13.00. Finally the dashed line
is a 5 Myr isochrone from Girardi et al. (2000).
In the HR diagram the number along the sequence indicate the star
masses. See text for additional details.}
\end{figure*}

\begin{figure*}
\centering
\includegraphics[width=16cm,height=16cm]{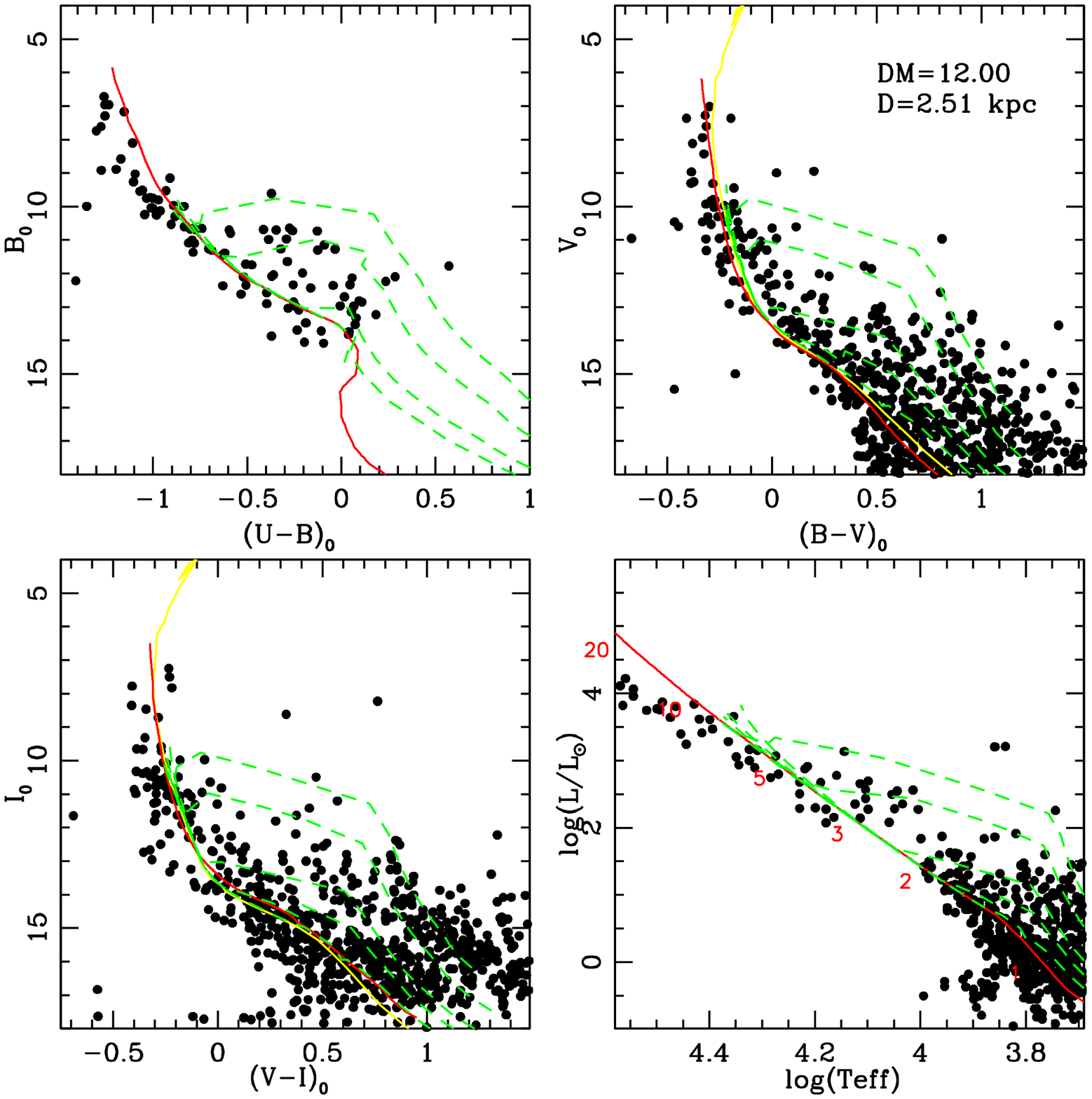}   
\caption{Same as Fig.~9, but for $(m-M)_0$=12. The poor agreement with the
data in the $B_0$-$(U-B)_0$ and HR diagrams for this distance modulus
is apparent.}
\end{figure*}

\subsection{Distances}
The clusters distances have been calculated by super-imposing
the observed points to an empirical Main Sequence (MS). Particular
care was taken to fit the upper part of the diagram, which is
populated by intermediate and high mass stars, for which we may
assume, due to the rapidity of their pre-MS evolution, that the Zero
Age MS corresponds to the observed MS.  In performing the fit we paid
attention to reproduce the bulk of the stars simultaneously in 3 CMDs
($B_o$ vs $(U-B)_o$, $V_o$ vs $(B-V)_o$, $I_o$ vs $(V-I)_o$) and in
the HR diagram, which are presented in Figs.~7, 8 and 9 for
Trumpler~14, Collinder~232 Trumpler~16, respectively.  This
strategy is mainly motivated by the almost vertical shape of the MS in
the $V_o$ vs $(B-V)_o$, which alone prevents reliable conclusions on
the distance of any star cluster, and takes full advantage of the more
favorable shape of the $U_o$ vs $(U-B)_o$ and HR diagrams.\\ Another
point to be emphasized is that we did not perform a membership
selection for all the clusters. This is due to the fact that proper
motions from Cudworth et al. (1993) are available only for stars
brighter than 15 mag in V, where the contamination of field stars is
less severe, and also to the fact that we are actually covering the
inner regions of the three clusters. This does not mean that we are
going to consider field star contamination ineffective. Only, we
believe that field star contamination does not alter our analysis and
conclusions significantly (but see the discussion below).  However, we
do cross-correlate our Trumpler~16 data with Cudworth et al. (1993) one,
in order to clean the upper part of the MS, which in the case of this
cluster is rather blurry. This fact not only helps us to better
constrain the cluster distance, but also to clarify whether the cluster
is actually somewhat older than the other two.\\

\noindent
{\bf Trumpler~14 (Fig.~7)} We find a good agreement between the
observed and the theoretical sequences in the four aforementioned
planes for Trumpler~14 by shifting the ZAMS by $(m-M)_o = 12.3\pm0.2$
(error by inspection), which implies a distance of $2.5\pm0.3$ kpc
from the Sun. We notice that this value is in perfect agreement with
the study by Vazquez et al. (1996, $(m-M)_o = 12.5\pm0.2$, error also
here by inspection), where a detailed analysis of the reddening has
been done as in our case, but with a different technique.  Tapia et
al. (2003) finally find $(m-M)_o = 12.23\pm0.67$, again in perfect
agreement with our findings.  This result gives us much confidence
when dealing with clusters (like Trumpler~16, see below) where the
field stars contamination is more severe.\\

\noindent
{\bf Collinder~232 (Fig.~8)} We find a good agreement between the
observed and the theoretical sequences in the four diagrams for
Collinder~232 by shifting the ZAMS by $(m-M)_o = 11.8\pm0.2$ (error by
inspection), which implies a distance of $2.3\pm0.3$ kpc from the
Sun. Basically, Collinder~232 is almost at the same distance as
Trumpler~14. The CMDs of Collinder~232 show the same features of those
of Trumpler~14 and 16 (see below), thus suggesting the possibility
that this cluster is probably a physical one.\\

\noindent
{\bf Trumpler~16 (Fig.~9 and 10)} The situation for Trumpler 16 is
somewhat more complicated, since the cluster is much more heavily
contaminated by field stars, and the upper part of the MS is rather
broad.  However, when proper motion members are considered, the
situation gets better.  Tapia et al. (2003) report a distance $(m-M)_o
= 12.02\pm0.57$, and place the cluster at the same distance of
Trumpler~14.  In the case of Trumpler~16, however, the traditional CMDs
$V_o$ vs $(B-V)_o$ (upper right panel) and $I_o$ vs $(V-I)_o$ (lower
left panel) do not help in finding a reliable value for the distance
modulus. In fact, the ZAMS in Fig.~9 and Fig~10 have been shifted by
$(m-M)_o$ = 13.00 (solid line) and 12.00 (dotted line), and do not
exhibit any real difference.  On the contrary, in the $B_o$ vs
$(U-B)_o$ CMD (upper left panel) and in the HR diagram (lower right
panel) they detach much more significantly, and only the larger
distance modulus ZAMS provides a good fit of the data (see Fig.~9 and
10). In conclusion, taking advantage of the large colour baseline, we
can reach a good fit by shifting the ZAMS by $(m-M)_o = 13.00\pm0.30$
(error by inspection), which in turn yields a distance of $3.98\pm0.5$
kpc. We note that this value is considerably larger than any previous
estimate of the distance of Trumpler~16.

\subsection{Ages and age spreads}
The age and age dispersion estimate is a cumbersome task. Our
theoretical tracks have been calculated by using the ATON2.0 code for
stellar evolution, a full description of which can be found in Ventura
et al. (1998). The pre-MS tracks are calculated starting from an
extremely cold structure ($\log T_C \sim 5.7$), and an evolutionary
status which takes place before the deuterium burning. This approach
can be adopted for the description of the early evolution of low mass
stars ($M\leq 1.5M_{\odot}$), but it is inadequate to determine the
age of more massive objects, since these latter complete deuterium
burning during the accretion phase, which is not taken into account
within our hydrostatic framework: for these stars we can only provide
an estimate of the time needed to reach the main sequence.  We
therefore set a minimum age for all the stars populating the MS, while
the analysis focused on the determination of the ages of the single
stars still in the pre-MS phase was limited to objects with mass
$M\leq 1.5M_{\odot}$.  As for massive stars, we are going to compare
their distribution with post-MS isochrone from Girardi et al. (2000).\\

\noindent
{\bf Trumpler~14 (Fig.~7)} Fig.7 shows the observed stars of Trumpler
14, along with our theoretical isochrones (Ventura et al. 1998, dashed
lines) corresponding to ages of 0.5,1,5,10 and $20 \times 10^6$ yrs
from the bottom to the top. The solid line is an empirical ZAMS,
whereas the long-dashed line is a 2 Myr  
post-MS isochrone
from Padova models (Girardi et al. 2000).
We note a well populated MS down to $M\sim
2M_{\odot}$, with no hints of stars leaving the MS.
In the lower part of the CMD and HRD 
we see that the pre-MS
population lies systematically 
rightward the pre-MS isochrone corresponding to an age of $2\times 10^7$ yrs,
which can be considered as the maximum age dispersion of the 
intermediate-mass stars in the
cluster. As for the age, since there is no clear indication of massive stars
in the act of leaving the ZAMS, we suggest  a very young age (less
than 2 Myr) for Trumpler~14, in perfect agreement with previous
suggestions (Vazquez et al 1996).\\

\noindent
{\bf Collinder~232 (Fig.~8)} The situation concerning Collinder 232 is
much better defined. As in the case of Trumpler~14, we note a
homogeneously populated MS down to masses $M\sim 2M_{\odot}$, and a
population of pre-MS stars in the lower part of the diagram well
detached from the theoretical MS. We have therefore indications from
photometry that Collinder 232 is indeed a physical group, and we find
that the dispersion of the ages is again within $\sim 2\times
10^7$yrs. As for the age, the same kind of comments as in the case of
Trumpler ~14 (see above) can be done.

\noindent
{\bf Trumpler~16 (Fig.~9)} The analysis for Trumpler 16 is much more
complex because the upper MS is larger. Since we are taking only
proper motion members into account, the width of the upper MS has to
be considered as due to the presence of massive stars out
of the MS, which in turn implies that this cluster is older than
Trumpler~14. To clarify this issue, in Fig.~9 we have drawn also a
post-MS isochrone from Girardi et al. (2000) for the age of 5 million
years, which provides a reasonable  fit to the data, in the
sense that there is the evidence that even somewhat less massive stars
are in the act of leaving the MS.\\ 
Besides, in the lower part of the
diagram it is not completely clear which stars effectively belong to
the cluster, so that the dispersion of the ages is extremely hard to
define. The proximity of stars in the
lower part of the diagram to the MS seems to indicate a further older
low mass stars population, but this conclusion is made very uncertain
by the tentative knowledge of the effective membership of the faintest
stars.\\

\section{Discussion and conclusions}

The main motivation of this study was to clarify the nature of the
star aggregate Collinder~232, i.e. whether this is a physical cluster
or not, and to investigate the relationship of the cluster with the
other two main clusters in the Carina spiral feature, namely
Trumpler~14 and 16.  We have addressed these issues by analyzing
homogeneous photometry in the optical passbands for all the
clusters. \\ The first step has been to study the extinction pattern.
In analogy with some previous investigations we find that all the
clusters are affected by absorption in quite a different way, and that
it is not possible - as in some previous studies- to adopt the same
reddening law for the entire Carina region. Actually, even assuming
the same absorption law within a given cluster could be already a
rather crude approximation.\\ The second step was to derive individual
reddening, luminosity and effective temperature for each star inside a
cluster. We derived these quantities by employing the method recently
developed by Romaniello et al. (2002).\\ Then we analyzed several CMDs
and the HR diagram and obtained estimates of the age, age spread and
distances.

\subsection{Main conclusions}
In Table~9 we summarize the main findings of this study.  The
extinction toward these clusters is highly patchy, a fact not always
properly taken into account in previous investigations. Moreover, the
analysis of the CMDs of Trumpler~16 reveals that this cluster is
significantly detached from the other two clusters and located further
away along the Carina spiral arm. This result apparently contradicts
previous findings.  All three clusters contain a substantial pre-MS
population, whose precise membership and consistency however is
hampered by field stars contamination.  In addition, we find for all
the clusters age spreads amounting at most at 20 million years and
Trumpler~16 seems to be older than Collinder~232 and Trumpler~14.

\subsection{Is Collinder~232 a physical aggregate?}
Tapia et al. (2003) performed star counts in the field of
Collinder~232, and concluded that there is no cluster there, simply
because the star density profile is almost flat and close to the mean
field star density in this region.  However, they do not consider the
appearance of the various CMDs, whose detailed scrutiny reveals that
we are facing a population of young stars, as in the case of
Trumpler~14 and 16. In other words, the shape of the stars
distribution in the CMD is that of a young stellar population, and
indeed (see Fig.~1), the main feature of Collinder~232 is a sparse
grouping of bright stars.  We cannot however firmly exclude that
Collinder~232 is just part of Trumpler~14 halo. In fact the eastern
side of Trumpler~14 (see Fig.~1) toward Collinder~232 is much less
obscured than the western part, where the Great Carina nebulosity is
optically very thick.  However it is not clear why we should see this
bright stars concentration only eastward of Trumpler~14, and not, for
instance, northward or southward.  The hypothesis of Collinder~232
being part of Trumpler~14 would also be somewhat corroborated by the
conclusion of Cudworth et al.  (1993) study, that the most probable
members of Trumpler~16 are enclosed with a circle 4 arcmin large,
which is depicted in Fig.~1, and therefore Collinder~232 is not
expected to be part of Trumpler~16. In this respect, Vazquez et
al. (1996) report for Trumpler~14 a radius of 2.5 arcmin, to small for
Collinder~232 being part of Trumpler~14.\\ We therefore propose the
possibility - which further studies should better investigate- that
Collinder~232 is a rather sparse, bright stars dominated, young open
cluster.

\begin{table*}
\tabcolsep 0.50cm
\caption{Estimates of the fundamental parameter of the clusters under investigations.}
\begin{tabular}{ccccccc}
\hline
\multicolumn{1}{c}{Cluster} &
\multicolumn{1}{c}{$E(B-V)$} &
\multicolumn{1}{c}{$(m-M)_o$} & 
\multicolumn{1}{c}{$Distance$} &
\multicolumn{1}{c}{$Age$} &
\multicolumn{1}{c}{$Age spread$}\\
\hline   
& mag & mag & kpc & Myr & Myr \\
\hline
Trumpler~16   & 0.61$\pm$0.15 & 13.00$\pm$0.30 & 3.9$\pm$0.5 & $\approx$ 5  & $\approx$ 20\\
Trumpler~14   & 0.57$\pm$0.12 & 12.00$\pm$0.20 & 2.5$\pm$0.3 & $\leq$ 2  & $\approx$ 20\\
Collinder~232 & 0.48$\pm$0.12 & 11.80$\pm$0.20 & 2.3$\pm$0.3 & $\leq$ 2  & $\approx$ 20\\ 
\hline
\end{tabular}
\end{table*}

\subsection{The Star Formation History in the $\eta$ Carina region}

According to Walborn (1995) and Megeath et al. (1996) SF is still
active in the Carina region. The analysis of Trumpler~14 and 16 lead
DeGioia-Eastwood et al.(2001) to conclude that intermediate-mass stars
started forming about 10 Myr ago, whereas high mass stars formed only
in the last 3 Myr.\\ Here we address a different issue, whether the SF
in this region has been sequential or not, following Feinstein (1995)
terminology.  We put together the results of our series of papers
(Carraro et al. 2001, Patat \& Carraro 2001, Carraro \& Patat 2001,
Carraro 2002, Baume et al. 2003 and the present one), where a
homogeneous data set has been presented and analyzed to derive in a
homogeneous fashion the ages of the young clusters in the Carina
region listed in Feinstein (1995, Table~1).\\

We notice that the youngest aggregates are Trumpler~14, Collinder~232
and Trumpler~16, which are located in the core of the region. A bit
further away, along the southern and northern extension of the arm,
there are NGC~3324 (same age as Trumpler~16), 
Trumpler~15 (6 Myr), Collinder~228 (8 Myr) and Bochum~11 (4 Myr), which
are also somewhat older.  NGC~3293 (10 Myr)
and NGC~3114 (300 Myr), located in a most
peripheral zone, are again somewhat older. Finally, Bochum~9 and 10
are probably not physical clusters, Collinder~234 seems to be part of
Trumpler~16, and VdB-Hagen~99 and Carraro~1 are not related with the
Carina spiral feature.\\
\noindent
In other word a clue emerges of a shallow age gradient along the spiral arm, 
which
seems to imply that SF started outside the $\eta$ Carina region
proceeding toward the core. This basically confirm the suggestions
made by Feinstein (1995).

\begin{acknowledgements}
The authors deeply thanks the anonymous referee for the her/his
encouraging and detailed report who helped a lot to improve on the
presentation of the paper.  G.C. acknowledges kind hospitality from
ESO and Rome Observatory, and very fruitful discussions with Gustavo
Luiz Baume and Guido Barbaro.  This study made use of Simbad and
WEBDA.
\end{acknowledgements}

{}

\end{document}